\documentclass{aa}  

\usepackage{graphicx}
\usepackage{txfonts}
\usepackage{lipsum}
\usepackage{subcaption}         % necessary for continued figures, example in section 3
\usepackage{lscape}             % to rotate a single page table, example in appendix.
\usepackage{placeins}           % useful with \FloatBarrier, to keep 
\newcommand{\htwo}{$\text{H$_2$}$}
\newcommand{\htwoo}{$\text{H$_2$O}$}

\newcommand{\ntwo}{$\text{N$_2$}$}

\newcommand{\cotwo}{$\text{CO$_2$}$}
\newcommand{\phthree}{$\text{PH$_3$}$}
\newcommand{\ashthree}{$\text{AsH$_3$}$}
\newcommand{\nhthree}{$\text{NH$_3$}$}
\newcommand{\chfour}{$\text{CH$_4$}$}
\newcommand{\gehfour}{$\text{GeH$_4$}$}
\newcommand{\hthreep}{$\text{H$_3^+$}$}                       

\begin{document}

   \title{Surveying exogenous species in Saturn with ALMA}

   \subtitle{II. Detecting and Mapping HCN}

%%%%%%%%%%%%%%%%%%%%%%%%%%%%%%%%%%%%%%%%
% Please do not include ORCIDs next to author names.
% Only ORCIDs authenticated by individual authors in EDP Sciences editorial system will be taken into account.
% ORCIDs included here will be removed.
%%%%%%%%%%%%%%%%%%%%%%%%%%%%%%%%%%%%%%%%

   \author{T. Fouchet\inst{1} \corrauth{thierry.fouchet@obspm.fr}
        \and T. Cavalié\inst{2,1}\email{thibault.cavalie@u-bordeaux.fr}
        \and R. Moreno\inst{1}\email{raphael.moreno@obspm.fr}
        \and S. Guerlet\inst{3,1}\email{sandrine.guerlet@lmd.ipsl.fr}
        \and E. Lellouch\inst{1}\email{emmanuel.lellouch@obspm.fr}
        \and D. Bardet\inst{1}\email{deborah.bardet@obspm.fr}
        }

   \institute{LIRA, Observatoire de Paris, Université PSL, CNRS, Sorbonne Université, Université Paris Cité, 5 place Jules Janssen, 92195 Meudon, France
             %\thanks{Shows the usage of elements in the author field}
            \and Univ. Bordeaux, CNRS, LAB, UMR 5804, F-33600 Pessac, France
            \and LMD/IPSL, CNRS, Sorbonne Université, École Polytechnique, Institut Polytechnique de Paris, École Normale Supérieure, Université PSL, 75005 Paris, France}

 %  \date{Received September 30, 20XX}

% \abstract{}{}{}{}{}
% 5 {} token are mandatory
 
  \abstract
   {Several external sources (the rings, the satellites, cometary impacts, interplanetary dust particles) supply material to the upper atmosphere of Saturn. The signature of this external influx has been documented on the ionospheric electronic and ionic densities, and in the stratosphere on oxygenated molecules (CO, CO$_2$, H$_2$O).}
   {We search for the chemical signature of the external sources in nitrogen-bearing species in Saturn's stratosphere.} 
   {We have targeted the HCN(4-3) rotational line and mapped its emission line at the planet limb with ALMA in 2018 with configuration C43-3, providing a synthetic beam of $0.61\arcsec\times0.50\arcsec$.}
   {HCN was detected at all observable latitudes from $20^{\circ}$S to $60^{\circ}$N. The HCN abundance was measured at two independent pressure levels, 0.1 and 1~hPa. The measured volume mixing ratios are latitudinally uniform within uncertainties at both pressure levels. The mean vmr is $(7\pm2)\times10^{-11}$ and $(4\pm1)\times10^{-12}$ at 0.1 and 1~hPa respectively. This corresponds to a mean column-density of $(4\pm1)\times10^{12}$~cm$^{-2}$, and a total nitrogen mass of $(3.7\pm0.9)\times10^{10}$~g.}
   {The steep vertical profile suggests that another loss mechanism, possibly sticking or heterogeneous chemistry on aerosols, is active besides photodissociation. The huge influx of material detected in the ring plane by Cassini left no detectable chemical signature on HCN. The detected HCN mass seems compatible with several sources: the \textit{ring rain} at mid-latitudes, Enceladus, or an old cometary impact. The detected HCN could also be a chemical product of internal \ntwo\ transported from the deep atmosphere up to the stratosphere.}


   \maketitle

%%%%%%%%%%%%%%%%%%%%%%%%%%%%%%%%%%%%%%%%%%%%%%%%%%%%%%%%%%%%%%
\nolinenumbers 
\section{Introduction}

Saturn is subject to a variety of distinct exogenous sources of material---the rings and the satellites, possible cometary impacts, and interplanetary dust particles \citep{Moses2017}---that can significantly influence the chemical composition of its atmosphere. While these sources have been primarily identified as delivering oxygen atoms in the form of water or carbon-bearing species such as CO and \cotwo\ \citep{Feuchtgruber1997}, they can also provide nitrogen atoms to Saturn's atmosphere.

Firstly, nitrogen has been clearly detected in the most recently identified and most prominent of all exogenous sources: the equatorial inflow of material from the rings into the atmosphere. Out of the 4,800 to 45,000 kg/s infalling mass flux dominated by oxygen and carbon, \citet{Waite2018}, using the Ion and Neutral Mass Spectrometer (INMS), estimated that $(2.4\pm0.5)\%$ of this inflow was in the form of \nhthree, $(20\pm3)\%$ in the form of CO or \ntwo, and $(37\pm5)\%$ in the form of heavy organics that may also contain nitrogen atoms. This initial analysis was later refined by \citet{Serigano2022}, who derived mean mass deposition rates of $(63\pm15)\times10^2$~kg/s for \ntwo, $(20\pm5)\times10^2$~kg/s for \nhthree, and $(19\pm9)\times10^2$~kg/s for HCN. Variations from orbit to orbit are present, by a factor two for \ntwo, three for HCN, and twenty for \nhthree. This flux is strongly localized within $\pm8^{\circ}$ of latitude from the equator and is thought to originate from a perturbation in the D68 ringlet that was spotted by Cassini in 2015 \citep{Hedman2019}.

Secondly, nitrogen-bearing species have been also clearly detected in the plumes of Enceladus. The moons's cryovolcanic activity releases 150--400 kg/s of material in the Kronian system, primarily in the form of \htwoo\ and \cotwo\ \citep{Hansen2006,Waite2006}. \citet{Waite2009} estimated that ammonia accounts for a fraction of $(8.2\pm0.2)\times10^{-3}$ of the plume composition, while \ntwo\ and HCN were not firmly detected with upper limit of $1.1\times10^{-2}$ and $7.4\times10^{-3}$, respectively. Using Cassini's Cosmic Dust Analyzer (CDA), \citet{Postberg2018} also identified heavy organic compounds, some of which contain nitrogen. Enceladus' plumes are thought to supply Saturn's E ring with particles and to feed the cold water torus detected in the gas phase by Herschel \citep{Hartogh2011}. A fraction of this material eventually precipitates into Saturn's atmosphere. \citet{Cassidy2010}, \citet{Hartogh2011}, and \citet{Moses2017} estimated that Enceladus provides $\sim2.5\times10^{26}$ (OH+\htwoo) per seconds (ie. $\sim6\times10^5$~cm$^{-2}$.s$^{-1}$), a flux consistent with the $(1\pm0.5)\times10^6$~cm$^{-2}$.s$^{-1}$ oxygen atom flux inferred from the ISO detection of Saturn's stratospheric water vapor \citep{Feuchtgruber1997,Moses2000}. Moreover, using Herschel mapping of Saturn, \citet{Cavalie2019} showed that Saturn's stratospheric water abundance peaks at the equator and decreases poleward, supporting the conclusion of \citet{Moses2017} that Enceladus dominates other external water sources. In addition, \citet{Donoghue2019} suggested that part of the material released by Enceladus in the magnetosphere, once ionized, may plunge along the magnetic field lines into Saturn's atmosphere. Hence, given the nitrogen-to-oxygen ratio in Enceladus' plumes, it can be expected that about 1\% of the $\sim10^{26}$~molec.s$^{-1}$ reaching Saturn from Enceladus could be in the form of nitrogen.

Finally, \citet{Moses2017} calculated that interplanetary dust particles (IDPs) deliver $\bigl(7.4^{+16}_{-5.1}\bigr)\times10^4$~O~atoms~cm$^{-2}$~s$^{-1}$, corresponding to a planetary-wide flux of $\bigl(3.4^{+7.3}_{-2.3}\bigr) \times 10^{25}$~O~atoms~s$^{-1}$. The elementary composition of IDPs at Saturn's heliocentric distance is not accurately known, but the progenitors of the IDPs in the outer Solar System are mostly cometary objects. Our knowledge of the composition of cometary objects is mostly based on observations of the coma. Compiling all reported observations, \citet{Bockelee-Morvan2017} reported that about $1\%$ of the cometary ices should be in the form of nitrogen-bearing species. Adopting this $1\%$ fraction for IDPs results in a global mass flux of $\bigl(7.9^{+17}_{-5.4}\bigr)$~g~s$^{-1}$ in the form of nitrogen atoms.

Nitrogen has not yet been directly identified in the other known exogenous sources, but there are strong reasons to believe that they also deliver nitrogen atoms, together with oxygen and carbon atoms. Among these sources, the rings contribute through an additional and well-characterized mechanism known as \textit{ring rain}. In this mechanism, dust sputtered from the rings becomes ionized and is transported along magnetic field lines into Saturn's upper atmosphere \citep{Northrop1982,Ip1983}. This influx was directly detected by the Cassini Cosmic Dust Analyzer (CDA) as grains, a few tens of nanometer in size, concentrated along field lines \citep{Hsu2018}. These authors estimated the associated mass flux at Saturn to lie between 100 and 360 kg/s. This value is consistent with indirect estimates based on the effects of the incoming ionized dust on Saturn's ionospheric electronic density \citep{Connerney1984} and on the \hthreep\ column density \citep{ODonoghue2013,Donoghue2019,Moore2015} at latitudes magnetically conjugated with the rings (23\degr--49\degr S and 32\degr--54\degr N). By modeling the \hthreep\ measured column density with ionospheric chemical model including a flux of ionized oxygen, \citet{Moore2015} and \citet{Donoghue2019} estimated the oxygen atom flux to be in the range of 2-200~kg/s. Since nitrogen-bearing species are present in dust originating from the rings according to INMS measurements \citep{Waite2018,Serigano2022}, it appears highly plausible that this \textit{ring rain} can also deliver nitrogen atoms to Saturn's mid-latitudes.

Unlike the other oxygen sources, which all originate within the Kronian system, \citet{Cavalie2009,Cavalie2010} proposed that the $(2.1\pm0.4)\times10^{15}$~g of CO they detected in Saturn’s atmosphere was supplied by a cometary impact that occurred $220\pm30$~years ago. This cometary impact is further supported by the model of \citet{Moses2017}, who showed that the ablation of grains originating from the rings or Enceladus, followed by the photochemical evolution of the released gases, could not account for the observed CO/\htwoo\ ratio. In contrast, they concurred that a recent cometary impact, combined with the continuous influx of water from Enceladus, can reproduce the observed CO/\htwoo\ ratio. Cometary impacts are known to produce and transport nitrogen-bearing species into the stratospheres of the Giant Planets. This process was directly observed during the SL9 impacts on Jupiter, with the production and transport of HCN and \nhthree, the latter being rapidly converted into HCN by photochemistry \citep{Marten1995,Moses1996,Bezard1997,Griffith1997,Lellouch2006,Cavalie2023}. In Neptune, recent studies have also showed convincingly that the stratospheric CO and HCN are most likely the remnants of a cometary impact that took place about two centuries ago \citep{Lellouch2005,Hesman2007,Luszcz-Cook2013}. If a cometary impact indeed introduced $(2.1\pm0.4)\times10^{15}$~g of CO in Saturn, it therefore seems highly probable that it also produced HCN, with a CO/HCN ratio that could be similar to the $\sim12$ value estimated for Jupiter or $\sim75$ for Neptune \citep{Lellouch2005,Lellouch2006}.

In addition to identified exogenous oxygen sources that may also provide nitrogen, the Saturnian system also hosts a major nitrogen-rich source: Titan and its thick \ntwo\ atmosphere. Models and \textit{in situ} measurements by Cassini have constrained the nitrogen escape from Titan's upper atmosphere. Using a 1D model constrained by Cassini measurements, \citet{Erkaev2021}  proposed a total flux of $\sim6\times10^{25}$~atoms.s$^{-1}$, combining all known escape mechanisms under the current solar UV flux. Assuming a homogeneous and isotropic escape, this rate corresponds to a nitrogen atom flux of $\sim4\times10^{22}$~atoms.s$^{-1}$ into Saturn's atmosphere.

Table~\ref{TabSources} presents the various nitrogen sources presented above, along with their measured or estimated planetary-averaged flux or planetary-integrated mass flux. 

Finally, we must not overlook that endogenous sources of nitrogen may exist in Saturn's atmosphere, where \nhthree\ dominates the nitrogen-bearing species. The \nhthree\ abundance in the stratosphere is quenched by condensation at the tropopause, where temperatures around 80~K act as a cold trap; therefore \nhthree\ should not reach detectable levels in the stratosphere. But nitrogen may also be present in the form of \ntwo, a gas that does not condense under any thermal conditions present in Saturn's atmosphere. Based on coupled thermochemical and eddy transport calculations, \citet{Fegley1985} predicted a \ntwo\ mixing ratio of $0.2-2\times10^{-6}$, depending on whether iron-catalyzed heterogeneous reactions are effective or not. More recent studies by \citet{Wang2016} and \citet{Cavalie2024} predicted a similar abundance of $\sim1.5\times10^{-6}$ and $(4\pm2)\times10^{-6}$ respectively. For Jupiter, \citet{Knizek2026} recently predicted that \ntwo\ transported to the upper atmosphere could be photolyzed to produce HCN with a volume mixing ratio (vmr) of $\sim10^{-8}$ in the upper stratosphere. On Saturn, the chemical model of \citet{Moses2023} also shows that the presence of \ntwo\ in the upper atmosphere can lead to the local formation of HCN. ALMA observations lent support to this production pathway: \citet{Cavalie2023} detected a local enhancement of HCN in Jupiter's southern auroral region at pressures smaller than 0.1 hPa, consistent with \textit{in situ} production.

With such a large variety of internal and external sources, it is somewhat surprising that no nitrogen-bearing species have yet been detected in Saturn’s stratosphere. Here, we report the first detection and mapping of the HCN (4-3) line on Saturn using ALMA. Section~\ref{SecObs} describes the observations, and Section~\ref{SecRadTran} the radiative transfer model and the inversion scheme, while the retrieved vertical and meridional profiles are presented in Section~\ref{SecResults}. These results are discussed in Section~\ref{SecDiscussion}, taking into account the derivation of the CO abundance profiles obtained from the same ALMA program and presented in a companion paper \citep{Bardet2026}.

\begin{table*}
\caption{Estimated nitrogen atom flux from known and expected external sources}
\label{TabSources}
\centering
\begin{tabular}{lccccr}
\hline\hline
Source & Planetary-averaged & Planetary-integrated & Occurrence & Localized & Comments \\
% \endfirsthead
% \caption{continued.}\\
% \hline\hline
% Source & Planetary-averaged & Planetary-integrated & Occurrence & Localized & Comments \\
% \hline
% \endhead
% \hline
% \endfoot
%%
 & flux (cm$^{-2}$s$^{-1}$) & mass flux (g/s) & &  \\ % table heading
 \hline
Ring Plane    & $(9.6\pm2.7)\times10^8$ & $(1.0\pm0.3)\times10^7$ & unknown & Sharp Equator & \citet{Serigano2022}\\
Ring rain  & (1.9--190)$\times10^3$ & 20--2000 & steady-state & Mid-latitude & $1\%$ of \citet{Donoghue2019}\\
Enceladus    & $\sim6\times10^3$ &  $\sim60$ & steady-state & Large Equator & $1\%$ of \citet{Hartogh2011}\\
Titan       &  $\sim90$    &  $\sim1$    & steady-state  & Planetary-wide   & \citet{Erkaev2021}\\
IDPs        & $\bigl(7.4^{+16}_{-5.1}\bigr)\times10^2$    &  $\bigl(7.9^{+17}_{-5.4}\bigr)$    & steady-state  & Planetary-wide   & $1\%$ of \citet{Moses2017}\\
Comet  &  & $(1.4\times0.3)\times10^{13}$~g & transient& Local+transport & \cite{Cavalie2010} \\
\hline
\end{tabular}
\end{table*}
%%%%%%%%%%%%%%%%%%%%%%%%%%%%%%%%%%%%%%%%%%%%%%%%%%%%%%%%%%%%%%
\section{Observations}
\label{SecObs}

We used the Atacama Large Millimeter/Submillimeter Array (ALMA) on 25 May 2018 to map Saturn’s atmospheric emission of HCN (4-3) at 354.5054759 GHz. The spectral setup was primarily configured to map the CO abundance and measure the winds through the Doppler shift of the CO (3-2) transition at 345~GHz \citep{Benmahi2022}, but also enabled a serendipitous search for HCN with a spectral resolution of 500~kHz. Hence, the observations (ALMA project 2017.1.00636.S, PI:T.\ Fouchet) were already presented in \citet{Benmahi2022} and \citet{Bardet2026}, showing the retrieved stratospheric zonal winds and the inferred CO vertical and meridional profiles, respectively.

The program was carried out with two scheduling blocks (SBs) obtained under excellent atmospheric conditions (precipitable water vapour of approximately 0.6 mm). Each SB included 36 minutes of on-source integration time. The first SB began at 04:31 UTC, at a central meridian longitude (CML) of 337\degr W (System III), and the second at 07:43 UTC (CML = 85\degr W). At the time of the observations, Saturn’s equatorial diameter was 18.04\arcsec\ and the sub-Earth latitude was 30.54\degr N. The longitudinal smearing during each SB was about 20\degr.

To cover the entire planetary disk emission, we performed a mosaic of seven pointings using the main array, configured in C43–2 with 43 antennas, yielding a synthetic beam of $ 0.61\arcsec \times 0.50\arcsec$ at the frequency of HCN (4-3). The corresponding spatial resolution on the disk of Saturn is approximately 5\degr\ in the equatorial and mid-latitude regions, and about 10\degr at northern polar latitudes.

\begin{figure}
        \centering
        \includegraphics[width=\hsize]{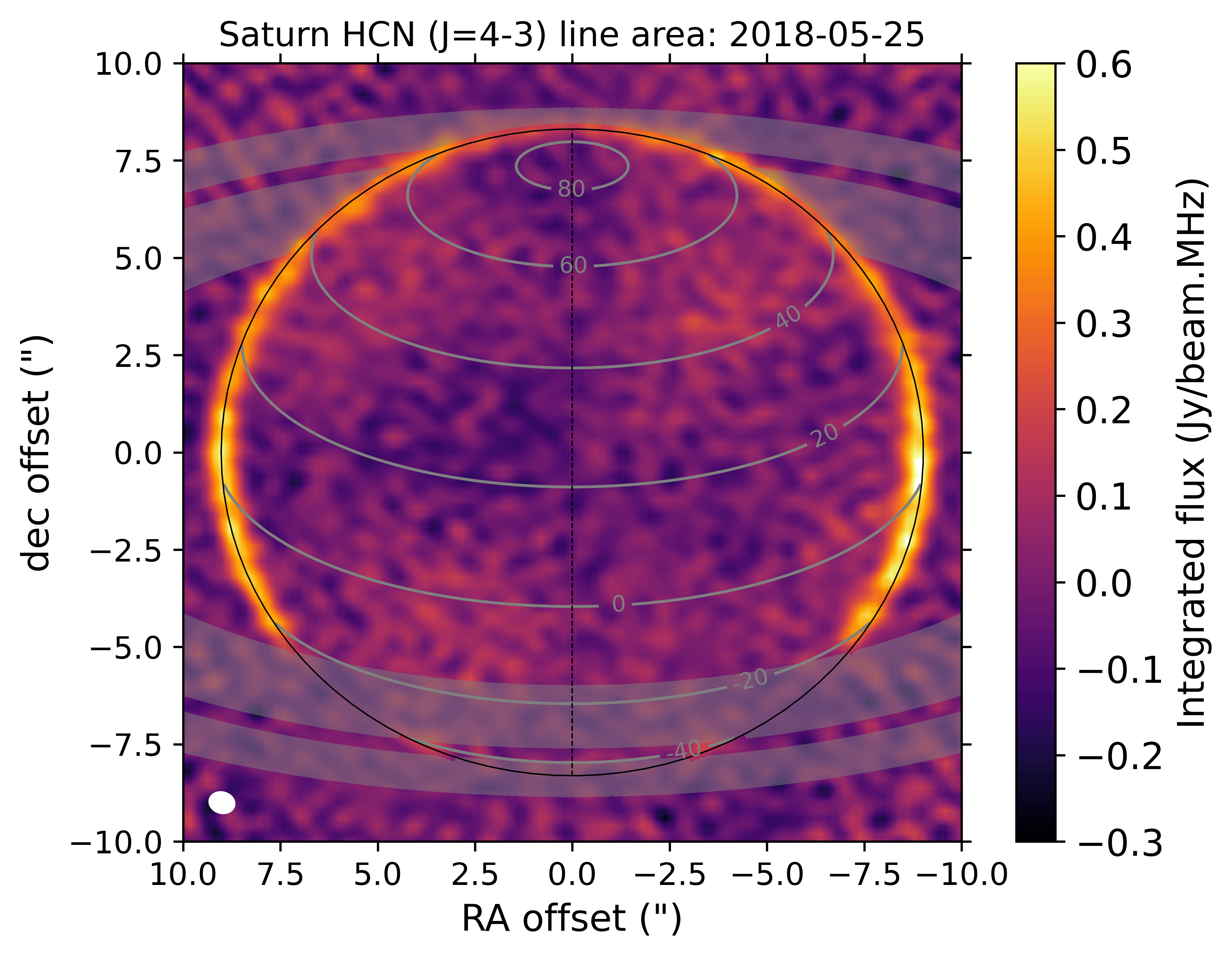}
        \caption{HCN (4-3) line area in Jy.MHz.beam$^{-1}$. The synthetic beam is highlighted as the white spot in the lower left corner.}
        \label{FigLineArea}
    \end{figure}

\begin{figure}
    \centering
    \includegraphics[width=\linewidth, clip=true, trim=50mm 50mm 60mm 105mm]{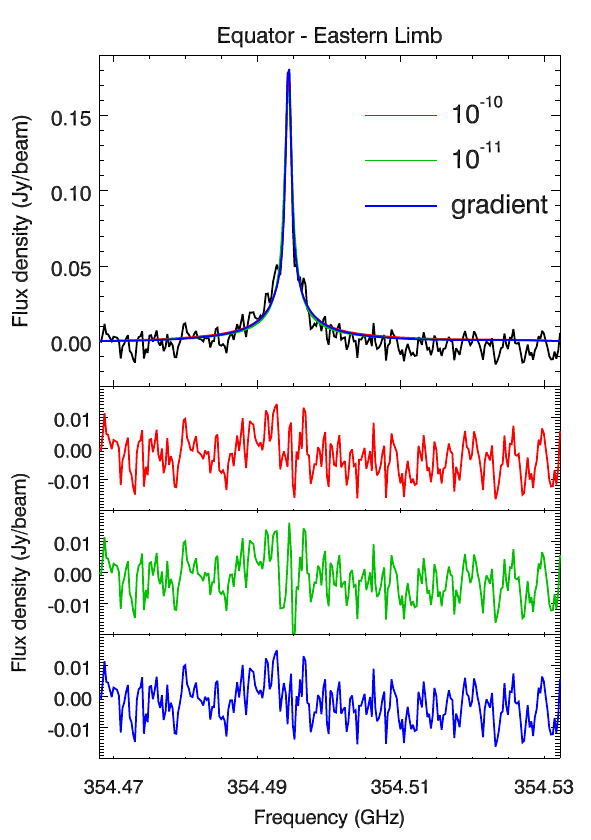}
    \caption{Comparison between the observed spectrum at the equator on the eastern limb (black line) and three synthetic spectra generated from the inverted profiles corresponding to the uniform $10^{-10}$ (red) \&\ $10^{-11}$ (green), and \textit{gradient} (blue) a priori profiles (upper panel). The lower panels present the residuals between the observed and synthetic spectra for the three inverted profiles (from top to bottom: using the $10^{-10}$, $10^{-11}$, and \textit{gradient} a priori profiles).}
    \label{FigSpectre}
\end{figure}

Data reduction was performed using the CASA software package (version 5.1.1.; \citealt{Muders2014}) with the standard ALMA pipeline. The visibilities were corrected for time-dependent atmospheric fluctuations in amplitude and phase through regular observations of the quasar J1832–2039. The radio-frequency (RF) bandpass was calibrated using the brighter quasars J1751+0939 and J1924–2914. The calibrated visibilities from each SB were then exported to the GILDAS package\footnote{https://www.iram.fr/IRAMFR/GILDAS/ and https://ascl.net/1305.010} \citep{Pety2005}, where the two SBs were averaged before producing the image cubes. Dirty images were cleaned using the Clark algorithm to generate the final clean images, after which Saturn was derotated so that its rotation axis aligned with the declination axis. Figure~\ref{FigLineArea} presents the resulting HCN line-area detected on both the eastern and western limbs of the planet, while Fig.~\ref{FigSpectre} presents the spectrum acquired at the equator on the eastern limb. No HCN absorption nor emission is detected on the disk of the planet.

%_____________________________________________________________
%                     Onecolumn continued float (place early!)
%-------------------------------------------------------------

\section{Radiative transfer code and inversion algorithm}
\label{SecRadTran}

To model our ALMA observations, we have used the forward model presented in \citet{Cavalie2019} and also used by \citet{Bardet2026}. This model is specifically designed to compute the emerging radiance at the limb by solving the radiative transfer equation on several thousands of line-of-sight within a beam. This allows to fully represent the large dependency of the radiance intensity with the tangent altitude. Emission is assumed to be at Local Thermodynamical Equilibrium (LTE) at all pressure levels in the atmosphere. The code further takes into account the thermal emission from the rings as a background emission source or a foreground absorption source, depending on the position of the line-of-sights.

We used the spectral line parameters from the JPL Molecular Spectroscopy Database \citep{Pickett1998}. We included the opacity due to collision-induced absorption caused by \htwo–\htwo, \htwo–He, and \htwo–\chfour\ pairs following the formalism of \citet{Borysow1985,Borysow1988} and \citet{Borysow1986}, adopting an helium volume mixing ratio of 11.8\% \citep{Conrath2000}, and of 0.47\% for methane \citep{Fletcher2009a}. The continuum opacity due to the broad wings of \nhthree\ and \phthree\ was also taken into account using the vertical distributions inferred by \cite{Davis1996} for \nhthree\ and \cite{Fletcher2009b} for \phthree, and broadening parameters proposed by \citet{Levy1993, Levy1994} for \phthree\ and  \cite{Fletcher2007} for \nhthree.

Retrieving the HCN abundance requires knowledge of Saturn’s stratospheric thermal structure at the epoch of the observations. Following \citet{Bardet2026}, we used the thermal structure derived from a combination of Cassini/CIRS measurements in the thermal infrared acquired in 2016 and 2017. In the equatorial region, the thermal structure was inferred between 10 and 0.01 hPa using data acquired in limb-viewing geometry \citep{Brown2024}. Outside the equatorial region, at latitudes south of 15\degr S and north of 10\degr N, the temperature structure was determined between 30 and 0.2 hPa from CIRS nadir spectra \citep{Fletcher2018}. Above the 0.2-hPa pressure level, the temperature profile was extrapolated as constant temperature with altitude. The adopted thermal structure is displayed in Fig.~3 of \citet{Bardet2026}.%\ref{FigThermalStructure}.

To retrieve the HCN vertical profile we used a constrained and regularized algorithm following the method introduced by \citet{Conrath2000}, and detailed in \citet{Fouchet2016} and \citet{Lellouch2017}. The inversion algorithm linearizes the radiance sensitivity with respect to the HCN vertical profile to obtain the Jacobian $K$, and uses a cost function that balances the fit to the observed spectra with the departure from the a priori profile and its vertical smoothness. This limits nonphysical oscillations in the retrieved profile at pressures probed by the dataset and unrealistic retrieved values at pressures not probed by our dataset. All our inversions were carried out with a vertical correlation length of $0.75$ times the atmospheric scale height.

\begin{figure*}
    \centering
    \includegraphics[width=\linewidth,clip=true, trim=20mm 55mm 00mm 160mm]{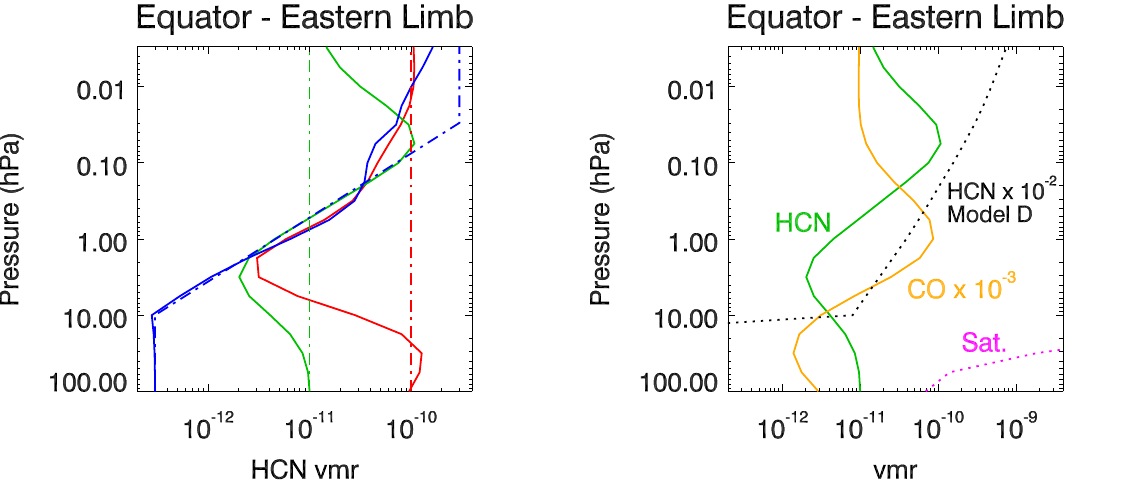}
    \caption{The HCN abundance vertical profiles (solid lines) inverted at the equator, eastern limb, using three different a priori profiles (dashed-dotted lines): uniform at vmr of $10^{-10}$ (red) and $10^{-11}$ (green), and the \textit{gradient} profile (blue) (left panel). Right Panel: The HCN abundance vertical profiles inverted at the equator, eastern limb using the $10^{-11}$ apriori (green) is compared with the HCN abundance from model D of \citet{Moses2023} (scaled by $10^{-2}$, black dotted), the HCN saturation abundance (magenta dotted, calculated using \citet{Fray2009}), and the CO abundance inverted by \citet{Bardet2026} from the same ALMA dataset (scaled by $10^{-3}$, solid orange).}
    \label{FigProfiles}
\end{figure*}

For the a priori profiles, we tested three different HCN vertical distributions in order to assess the pressure range of maximum sensitivity of our measurements, and to evaluate the influence of unconstrained abundances outside the sounded region on the retrieved volume mixing ratio within it. As shown in Fig.~\ref{FigProfiles}, two of the a priori profiles assume vertically uniform HCN abundances, with vmr of $10^{-10}$ and $10^{-11}$, respectively. The third one, intended to represent an external flux, is vertically uniform above 0.03 hPa and below 3 hPa, with vmr values of $3\times10^{-9}$ and $3\times10^{-12}$, respectively, and decreases linearly in log-log space between these two pressure levels. In the following, we will labeled this a priori profiles as \textit{gradient}.

\section{Results}
\label{SecResults}

The HCN inverted vertical profiles exhibit qualitatively similar characteristics at all beam positions on both limbs. These characteristics are illustrated in the left panel of Fig.~\ref{FigProfiles} for the spectrum acquired at the eastern limb of the equator. The corresponding synthetic spectra and residuals are shown in Fig.~\ref{FigSpectre}. The three inverted profiles yield consistent HCN abundances between 0.1 and 2~hPa, hence defining the pressure range over which our dataset is sensitive. Outside this range, at both lower and higher pressures, the profiles relax toward their respective a priori profiles. This sensitivity range is confirmed by $\sim2$ degrees of freedom and the averaging kernels (not shown, see \citet{Fouchet2016,Lellouch2017} for a definition of the averaging kernels) indicating that the HCN abundance can be independently measured at two different pressure levels. At each beam position, the inverted HCN vertical profile show a steep vertical gradient, with the abundance decreasing by roughly one order of magnitude per decade in pressure between 0.1 and 2~hPa (i.e.\ roughly constant HCN number density).

\begin{figure}[t]
    \centering
        \includegraphics[width=1\linewidth,clip=true,trim=30mm 50mm 40mm 143mm]{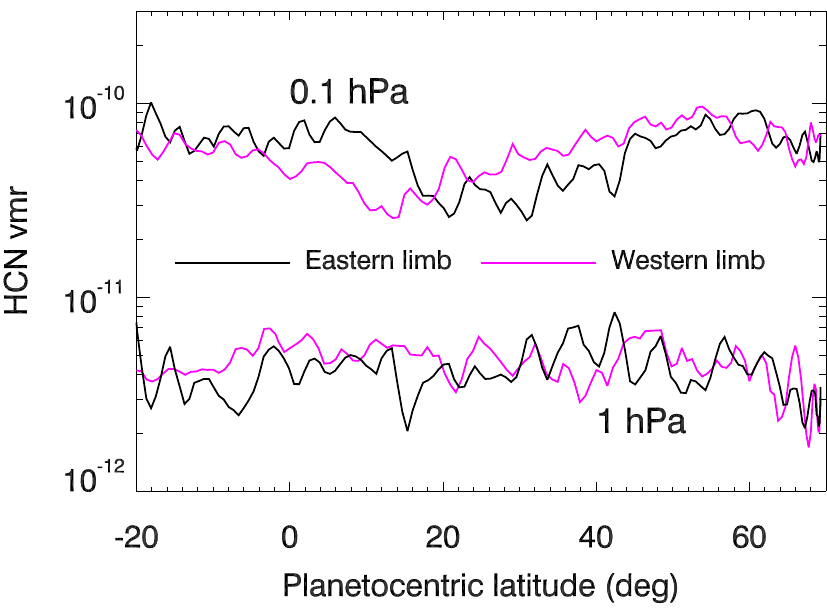}
        \includegraphics[width=1\linewidth,clip=true,trim=30mm 50mm 40mm 143mm]{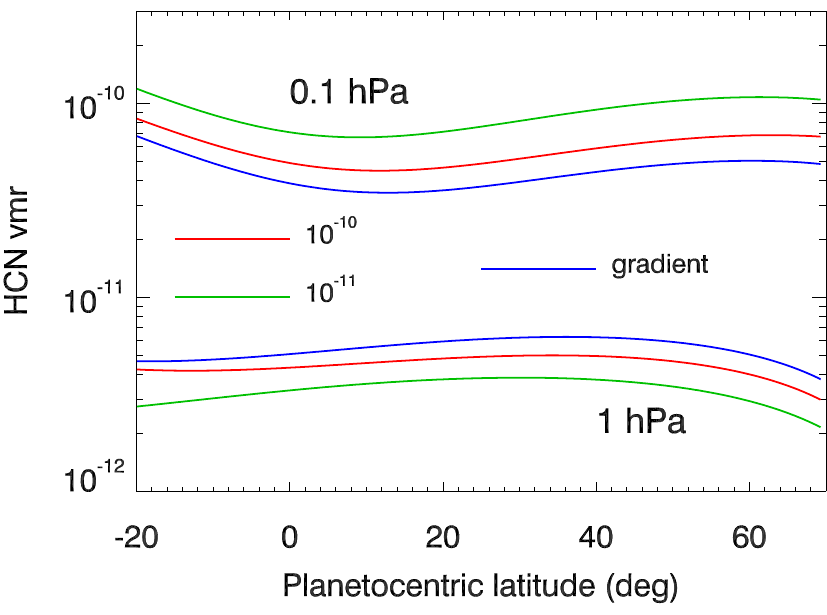}
    \caption{Upper Panel: Meridional profiles of HCN vmr retrieved at the eastern (black) and western (magenta) limbs at 0.1 and 1~hPa using the uniform a priori profile at $10^{-10}$. These profiles include measurements from all beam positions. Lower Panel: Smoothed meridional profiles at 0.1 and 1~hPa retrieved using the uniform a priori profiles of $10^{-10}$ (red) \&\ $10^{-11}$ (green), and the \textit{gradient} profile (blue).}
    \label{FigMeridional}
\end{figure}

Consistently, with the $\sim2$ degrees of freedom and the pressure sensitivity range of our inversion, Fig.~\ref{FigMeridional} presents the retrieved HCN vmr meridional profiles at two pressure levels, 0.1 and 1~hPa. For clarity, the upper panel shows the vmr retrieved at all beam positions on the eastern and western limbs, using only the $10^{-10}$ a priori profile. The abundances retrieved on the eastern and western limbs at similar latitudes agree very well, within $\pm 1.5\times10^{-11}$ at 0.1~hPa and $\pm 1\times10^{-12}$ at 1~hPa (i.e.\ within $\pm20\%$ in relative proportion). The absence of any sharp latitudinal features in the retrieved raw meridional profiles motivated us to present, in the lower panel of Fig.~\ref{FigMeridional}, latitudinally smoothed profiles obtained using the three different a priori profiles. The smoothing was performed by fitting the combined eastern and western raw meridional profiles with third-order polynomials, improving the clarity and distinction between the resulting three meridional profiles. Two main conclusions can be drawn from these profiles. First, the uncertainty in the retrieved HCN abundance is affected by the choice of the a priori profiles by the same amount as by the measurement noise. At 0.1~hPa, the meridional profiles differ systematically by $\pm2\times10^{-11}$, and by $\pm 1\times10^{-12}$ at 1~hPa. Second, the inferred meridional distributions are nearly uniform with latitude at both pressure levels, with a mean HCN vmr of $(7\pm2)\times10^{-11}$ at 0.1~hPa and $(4\pm1)\times10^{-12}$ at 1~hPa. The only notable deviations from this uniformity are a possible increase in abundance southward of the equator at 0.1~hPa, and a possible decrease northward of 50\degr N at 1~hPa. %In both cases, these variations tend to steepen the inferred vertical gradient of HCN at .

In terms of column-density, this corresponds to a mean value of $(4\pm1)\times10^{12}$~cm$^{-2}$ between the 2 and 0.05-hPa pressure levels. In terms of total nitrogen mass, assuming symmetry between the two hemispheres, our observations yield a total mass of nitrogen atoms contained in HCN of $(3.7\pm0.9)\times10^7$~kg between the equator and the 60\degr\ latitude circles, which encompass 85\%\ of the planet's surface.

\section{Discussion}
\label{SecDiscussion}

\subsection{The equatorial influx from the rings?}

As we will now show, our retrieved HCN abundances do not support a steady-state, equatorially centered influx of the magnitude measured by Cassini instruments during the final proximal orbits \citep{Waite2018,Serigano2022}. This conclusion is supported both by the absolute HCN abundances we measure and by their meridional variations.

\citet{Moses2023} investigated the impact of such a flux on stratospheric chemistry using a photochemical model, exploring a range of scenarios that account for variability in the measured flux from orbit to orbit, uncertainties in the dust-to-gas ratio in INMS measurements, and the possibility for meridional redistribution prior to atmospheric entry. In their lowest-flux scenario (case D), they assumed that only the most volatile molecules (\ntwo\ for nitrogen) enters the atmosphere in gaseous form, the other molecules being trapped in dust that never ablates, and divided the flux measured for Rev.~293 by \citet{Serigano2022} by a factor of 10 to account for meridional spreading. Still, this flux of $2.14\times10^8$~cm$^{-2}$.s$^{-1}$ results in predicted HCN abundances of $1.5\times10^{-8}$ at 0.1~hPa and $5\times10^{-9}$ at 1~hPa. These values are approximately 3 orders of magnitude higher than the mean abundances derived from our observations at the same pressure levels.

To assess whether the meridional variations are compatible with a strong equatorial influx, we estimate the maximum equatorial influx compatible with our inferred latitudinal profiles. The meridional flux is given by
\begin{equation}
    \Phi_y =  - K_{yy}n\frac{\partial q}{R_S\partial\theta}
\end{equation}
where $q$ is the HCN vmr, $\theta$ the latitude, $R_S$ Saturn's radius, $n$ the number density and $K_{yy}$ the horizontal eddy mixing coefficient. For this coefficient, we adopt the value derived by \cite{Moreno2003} for Jupiter from the spreading of SL9-induced species, $Kyy=2\times10^{11}$~cm$^2$.s$^{-1}$. Given the uncertainties in our measurements, the abundance latitudinal difference between the equator and 60\degr N must remain smaller than $4\times10^{-11}$ and $4\times10^{-12}$ at 0.1 and 1~hPa, respectively. These limits imply a horizontal flux lower than $6.5\times10^6$~cm$^{-2}$.s$^{-1}$ at the two pressure levels. Further assuming that this flux is zonally and vertically uniform over the 2.3 scale heights spanning the 0.1-1~hPa pressure range, this corresponds to a maximum flux of $3.6~\times10^{24}$ nitrogen atoms per second, or 80~g.s$^{-1}$ of \ntwo\ molecules. If this horizontal divergence is compensated by a downward transport at steady state, the maximum equatorial influx is about five orders of magnitude lower than the \ntwo\ influx measured by \citet{Serigano2022}, $(6.3\pm1.5)\times10^3$~kg/s, or the total nitrogen influx of $\sim10^4$~kg/s when all nitrogen-bearing species are considered. Of course, our estimate is derived from an equation that assumes that no loss nor production takes place in the modeled region, hence that all the HCN is introduced in the stratosphere at the equator and only destroyed at latitudes higher than 60\degr. The existence of local production and destruction pathways may alter our estimate, but the production and destruction rates would need to be unrealistically large to be compatible with a quasi-uniform meridional profile. We will address the question of the HCN loss further in a few paragraphs. We further note that two other exogenous species do not exhibit any signature of a strong equatorial influx, CO and \cotwo. For CO, the meridional profiles obtained by \citet{Bardet2026}, using the same ALMA dataset as used in this study, do not show any meridional gradient larger than $3\times10^{-8}$ vmr between the equator and 60\degr N. For \cotwo, \citet{Fletcher2023}, using JWST/MIRI spectra obtained in November 2022, found that the \cotwo\ abundance is meridionaly uniform within relative variations of $\pm10\%$.

Hence, both the absolute HCN abundance and meridional distribution inferred from our measurements are highly incompatible with a steady-state gaseous influx at the magnitude measured by Cassini, consistently with other molecular tracers of exogenous material. However, could our dataset and measurements be compatible with a transient flux, lasting at least over the six-month period during which Cassini's proximal orbits probed the region between the rings and the planet? To assess this possibility, we first note that our determination of the total nitrogen mass present in Saturn's atmosphere in the form of HCN, $(3.7\pm0.9)\times10^7$~kg, corresponds only to about 6,000~seconds of precipitation for a mean \ntwo\ influx of $6.3\times10^3$~kg/s, and about 3,600~s for all nitrogen-bearing species average influx of $1\times10^4$~kg/s. Such a duration is extremely short compared to the total duration of the proximal orbits ($\sim1.3\times10^7$~s), even more so when compared to the 2-3 years that elapsed between the D68 ringlet perturbation, hypothesized by \citet{Hedman2019} to be the source of the equatorial flux, and the proximal orbits. And it cannot be explained by the temporal variations of the flux observed from orbit to orbit, as these variations are within a factor of two for \ntwo\ and three for all nitrogen-bearing species. Alternatively, this $2000$-s timescale could be the signature not of the actual duration of precipitation, but rather the HCN chemical lifetime in Saturn's atmosphere. Such a short lifetime, corresponding to an enormously large loss rate, seems highly unrealistic, as discussed in the following paragraphs.

\begin{figure}[h]
    \centering
    \includegraphics[width=\linewidth, clip=true, trim=50mm 110mm 60mm 105mm]{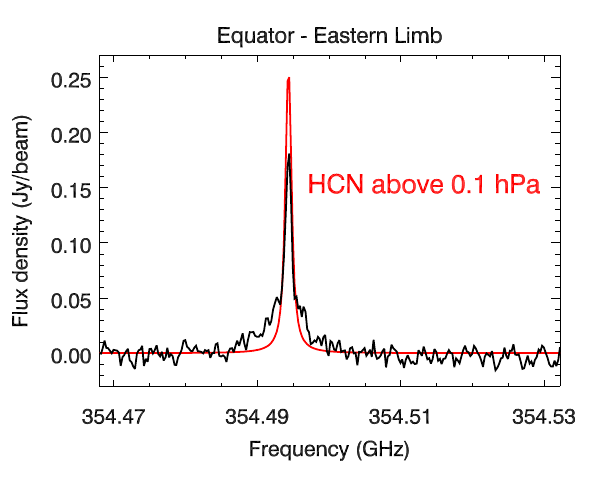}
    \caption{Comparison between the observed spectrum at the equator on the eastern limb (black line) and a synthetic spectrum (red) generated for an HCN column-density of $4\times10^{12}$~cm$^{-2}$ located exclusively above the 0.1-hPa pressure level with a vertical homogeneous distribution.}
    \label{FigSpectreUpper}
\end{figure}

If the large influx monitored by Cassini resulted from a recent transient event, the precipitating gas may not yet have diffused down to the pressure sensitivity range of our observations. Our ALMA data were obtained in May 2018, 8 months after Cassini's final plunge into Saturn's atmosphere, 13 months after the beginning of the proximal orbits, and approximately 4 years after the D68 perturbation. \citet{Moses2023} studied the temporal evolution of the HCN vertical profile following a transient ring influx diffusing down through the upper atmosphere. Their model indicates that, after 4 months, the influx should have crossed the homopause level and reached the upper stratosphere, and that the diffusion front should reach the 0.1–0.01 hPa region after 3 years. These atmospheric levels are located above the vertical sensitivity range of our observations but still contribute to the observed emission line. Hence, to confirm that our observations are incompatible with a significant fraction of the HCN column density residing above the 0.1-hPa pressure level, we have calculated a synthetic spectrum for an HCN column density of $4\times10^{12}$~cm$^{-2}$ residing only above the 0.1-hPa pressure level, a column density similar to the mean value retrieved between the 2 and 0.05-hPa pressure levels (Sec.\ref{SecResults}). This synthetic spectrum is compared to the spectrum observed at the equator and eastern limb in Fig.~\ref{FigSpectreUpper}. The integrated line-area obtained with this profile is similar to that obtained with the inversion process (Fig.~\ref{FigSpectre}) because at the low abundances considered the HCN(4-3) line is optically thin, and because Saturn's stratospheric temperature is nearly vertically uniform up to the homopause level at about 140--150K \citep{Brown2021}. Under these conditions, the HCN(4-3) line area is directly proportional to the HCN column density. However, the line profile calculated for an HCN column density restricted to the upper stratosphere is narrower than that observed and inverted, lacking some pressure broadening at pressure levels deeper than 0.1-hPa. We can therefore rule out the possibility that our analysis missed a significant amount of material injected into the upper stratosphere a few months or years prior to our observations. We note that a similar conclusion was reached by \citet{Moses2023}, who showed that the infrared signatures of HCN and other species should have been detected in Cassini/CIRS spectra if the ring influx had been converted into gaseous species in Saturn's atmosphere.

Overall, our HCN detection and mapping support the conclusion reached by \citet{Moses2023} that most of the incoming material, including \ntwo, must have entered the atmosphere as small dust grains that did not ablate during atmospheric entry.

\subsection{An additional HCN loss mechanism}

Before investigating which external sources could account for the presence of HCN in Saturn's stratosphere, we must first assess its loss rates. As stated in \citet{Moses2023}, HCN is relatively stable under Saturn's stratospheric conditions. To quantify this stability, we calculate the photochemical lifetime of HCN with a simple atmospheric model. The HCN UV cross-sections are taken from \cite{Nuth1982} and \cite{Lee1980}, and the CH$_4$ cross-sections from \cite{Backx1975}, \cite{Lee1983}, and \cite{Mount1977}. Using the solar reference spectrum from \cite{Thuillier2004}, we derive an HCN photolysis rate of $3.5\times10^{-7}$~s$^{-1}$ at the top of the atmosphere. At 0.1~hPa, taking into account self- and \chfour-shielding, this rate implies an HCN photolysis timescale of about 80 terrestrial years, and $\sim150$ years at 1~hPa. The actual net loss timescale is likely even longer, as underlined by \citet{Moses2023}, since the photolysis products are efficiently recycled into HCN. Consistently, the HCN vertical gradient modeled by \citet{Moses2023} is rather weak, with HCN decreasing by a factor of 2--3 between 0.1 and 1~hPa (right panel of Fig.~\ref{FigProfiles}). 

This is in striking contrast with our measurements, which indicate a decrease by a factor of $17.5^{+12.5}_{-7.5}$ between 0.1 and 1~hPa, and strongly suggest that an additional loss mechanism is not accounted for in current photochemical models. As shown in the right panel of Fig.~\ref{FigProfiles}, this sharp decrease is not due to saturation. A similar discrepancy between photochemical models and retrieved HCN profiles has been reported in Titan’s lower stratosphere, where the observed HCN vertical gradient is steeper than predicted by photochemical models \citep{Vinatier2007}. A comparable situation is also currently monitored in Jupiter's stratosphere, where the HCN produced by the SL9 impacts is efficiently destroyed in specific regions: the lower stratosphere of the polar regions \citep{Lellouch2006,Cavalie2023}. The localization of this HCN sink led \citet{Lara1999}, \citet{Vinatier2007}, and \citet{Cavalie2023} to propose that heterogeneous chemistry on stratospheric organic aerosols dominates the HCN loss rates in Jupiter and Titan. This hypothesis is supported by laboratory experiments demonstrating that HCN effectively binds to analogs of Titan's aerosols \citep{Perrin2021,Perrin2025}.

Organic aerosols are also present in Saturn's stratosphere. Their presence is evidenced by Cassini ISS limb images analyzed by \citet{Sanchez-Lavega2020}, who found that aerosols are primarily located between the 500 and 10~hPa pressure levels, with an exponential decrease in number density with altitude characterized by a scale height of H=19--21~km. In addition, an extended aerosol layer reaches an altitude of up to 340~km (above the $10^3$-hPa pressure level), at pressures as low as 0.4~hPa. The aerosols number density retrieved in Saturn's atmosphere around the 1~hPa pressure level, $\sim10^{-2}$~cm$^{-3}$ \citep{Sanchez-Lavega2020}, is quite comparable with that inferred for Jupiter \citep{Wong2003}. It is therefore highly plausible that the same loss mechanism, involving heterogeneous chemistry, is also active in Saturn's stratosphere.

We further note that our inverted HCN vertical profile is in sharp contrast with that inferred for CO by \citet{Bardet2026} (Fig.~\ref{FigProfiles}). In fact, within the 0.1--1~hPa pressure range, HCN and CO exhibit opposite vertical gradients, with HCN vmr decreasing and CO vmr increasing with pressure. This difference rules out possible alternative explanations for the steep HCN vertical gradient (weaker vertical diffusion than assumed, diffusion front of a recent comet impact), and strengthens our conclusion that HCN suffers an additional loss in the deep stratosphere.

\subsection{A permanent, planetary wide external source?}

The presence of this additional loss mechanism complicates the identification of the nitrogen exogenous source. Indeed, the HCN loss rate on aerosols is poorly constrained and depends on several uncertain microphysical parameters, including sticking coefficients, effective surface area, reaction rates, etc... This situation makes it more difficult to estimate the external flux required to explain the observed HCN distribution. However, an order-of-magnitude estimate can still be derived under steady-state conditions. To do so, we assume that most of the loss occurs at pressure levels deeper than those probed by our observations. This assumption is supported by the fact that aerosols are more abundant at lower altitudes \citep{Sanchez-Lavega2020}. Under these conditions, the descending flux implied by our vertical HCN profile is given by
\begin{equation}
    \Phi_z = - K_{zz}\,n\,\frac{dq}{dz}.
\end{equation}
where $K_{zz}$ the vertical eddy mixing coefficient and $z$ the altitude. Adopting the $K_{zz}$ value of $10^5$~cm$^2$s$^{-1}$ at 1~hPa that reproduces the observed ethane and acetylene profiles in \citet{Moses2023} photochemical model, we obtain an HCN downward flux of $(7.3\pm2.6)\times10^3$~cm$^{-2}$s$^{-1}$. Integrated over the surface of Saturn's, this corresponds to a planetary-wide mass flux of $(80\pm30)$~g/s in nitrogen atoms.

These values are significantly larger than those estimated for nitrogen originating from Titan's atmospheric escape (see Table~\ref{TabSources}), and also larger than what IDPs are expected to deliver \citep{Moses2017}. This indicates that a stronger source dominates the global nitrogen influx. At the other extreme, our estimates again rule out a significant conversion of the nitrogen contained in the equatorial ring flux detected by Cassini into atmospheric gases, even if the incoming material were efficiently redistributed across the entire planetary surface prior to atmospheric entry. If this flux were a permanent source, at most $\sim10^{-5}$ of the incoming mass would vaporize and contribute to the atmospheric chemical composition. Among the two remaining steady-state external sources, our mass flux estimates are consistent with both an Enceladus origin and the \textit{Ring rain} scenario. However, the nitrogen fraction in these sources is entirely unconstrained and, given the assumptions underlying our mass flux estimates, it would be an over-interpretation to favor one source over the other based on this analysis alone.

Nevertheless, although both sources are compatible with our estimate of the global mass flux, they are expected to be latitudinally localized on Saturn’s disk. The \textit{Ring rain} is observed at specific latitudes, magnetically conjugated with the rings and located between 40\degr\ and 50\degr\ in the northern hemisphere \citep{Donoghue2019}, whereas the Enceladus flux was modeled by \citet{Cavalie2019} as a gaussian centered on the equator with a latitudinal FWHM of $\pm25$\degr. Given the strong HCN loss rate implied by the observed steep vertical gradient, homogenizing the meridional HCN distribution to the observed uniformity would require highly efficient horizontal mixing. We note that a balanced combination of both sources could yield a meridional distribution close to uniform.

\subsection{An internal source?}

In this respect, an internal source of nitrogen may seem also promising in yielding a uniform meridional distribution, as convection is ubiquitous on Saturn. In the shallower atmosphere, the thermodynamic equilibrium abundance of HCN is extremely low, but it increases with pressure along Saturn's adiabat. If vertical mixing is more efficient than chemical quenching, a significant amount of HCN could be transported to the upper troposphere and stratosphere. Early work by \citet{Fegley1985} predicted an HCN vmr in the observable Saturnian atmosphere ranging from $10^{-12}$ to $10^{-7}$, whereas a more recent work by \citet{Wang2016} predicts a much lower value of $\sim2\times10^{-15}$. However, the presence of a substantial internal HCN source would give a vertical profile that decreases with altitude in the vertical region where loss on aerosols occurs and becomes uniform above this region, i.e., above the 1-hPa pressure level. Since such a profile is incompatible with our observations, a direct internal source of HCN appears highly unlikely, consistent with current thermodynamic and kinetic model for Saturn \citep{Wang2016}.

However, stratospheric HCN could also be produced indirectly from another disequilibrium gas, \ntwo. \citet{Wang2016} predict a \ntwo\ vmr of $\sim1.5\times10^{-6}$ in Saturn's observable atmosphere, while \cite{Knizek2026} predict an abundance of 11~ppm for Jupiter. In the upper stratosphere, this \ntwo\ could be photolysed or destroyed by ion-neutral reactions to form HCN \citep{Moses2023}. For Jupiter, \citet{Knizek2026} predicted a stratospheric HCN abundance of $1.5\times10^{-10}$ at 0.1~hPa from such processes, while \citet{Cavalie2023} reported HCN production in the Jovian auroral regions that they attributed to \ntwo\ ion-neutral reactions. Moreover, this formation pathway would be qualitatively consistent with our observed vertical profile, as HCN would be produced in the upper stratosphere and lower ionosphere. To our knowledge, no quantitative predictions of HCN production from this mechanism exist for Saturn. However, we can estimate its effect using the scenarios of \citet{Moses2023} in which only the most volatile species, \ntwo\ for the nitrogen-bearing species, vaporize in Saturn's atmosphere. Under these conditions, these authors predict an HCN-to-\ntwo\ ratio of $\sim10^{-2}$ from the destruction of \ntwo. Applying this ratio to our measured HCN vmr of $\sim10^{-10}$ at the 0.1~hPa pressure level yields a \ntwo\ vmr of $\sim10^{-8}$, an abundance about two orders of magnitude smaller than the prediction of \citet{Wang2016}. This analysis suggests that the chemical models of \citet{Wang2016} and \citet{Moses2023} may not be fully reconciled, with at least one potentially overestimating either the disequilibrium abundance of \ntwo\ or the production rate of HCN from \ntwo. Regarding our objective of identifying the source of HCN observed in Saturn's stratosphere, we cannot rule out the possibility that it is indirectly supplied by an internal source.

Whether such an indirect internal source is compatible with a uniform meridional profile of HCN remains an open question. Using a parameterization of the tropospheric eddy diffusion coefficient based on laboratory studies dedicated to turbulent rotating convection, \citet{Wang2015} predicted that the vertical transport efficiency decreases from the Equator towards the Polar Regions on Giant Planets. Using this parameterization, they predicted a meridional decrease for the disequilibrium species CO and \phthree\ by a factor of three on both Jupiter and Saturn. However, their predictions are not supported for Jupiter by the observations of \citet{Drossart1990} and \citet{Giles2017}, who found that the \phthree\ and \ashthree\ abundances increase poleward, whereas that of \gehfour\ remains more or less uniform. We further note that \ntwo, a very stable species in Saturn's upper troposphere and lower stratosphere, could be efficiently homogenized horizontally while being transported upward across the tropopause. Given all these uncertainties and open questions, it is difficult to assess whether an internal source contributes significantly to the observed stratospheric HCN.

\subsection{A cometary origin?}

Finally, we need to discuss the plausibility of a cometary impact providing the nitrogen atoms required to produce the HCN observed in Saturn's stratosphere. In this scenario, proposed by \citet{Cavalie2010} and \citet{Moses2017} to explain the presence of CO of \cotwo\ in Saturn's stratosphere, disequilibrium species are produced in the impact fireball and its following splashback, and deposited near the 0.1~hPa pressure level. Then, as observed on Jupiter \citep{Moreno2003,Griffith2004,Lellouch2006,Benmahi2020,Cavalie2023} and Neptune \citep{Lellouch2005,Hesman2007}, the products diffuse downward and horizontally, leading to a uniform distribution from pole to pole, as seen for CO on Jupiter \citet{Cavalie2023}, and to a vertically homogeneous profile above the downward diffusive front. Such a scenario for the origins of HCN is consistent with our observed vertical profile, providing that a strong loss mechanism operates in the lower stratosphere. In fact, our inferred profile closely matches that reported by \citet{Cavalie2023} in the polar regions of Jupiter, where the loss mechanism on aerosols is the most efficient. This scenario is also consistent with the inferred uniform meridional distribution since the impact on Saturn is estimated as $220\pm30$~years old by \citet{Cavalie2010}, a period much longer than the estimated horizontal mixing timescale.

Since the loss rate on aerosols is poorly constrained and very difficult to model, we consider it inappropriate to extrapolate the observed HCN mass load to its initial post-impact value and assess its plausibility. Instead, we rather adopt a reverse approach and estimate the HCN lifetime by assuming an initial mass load. Following the work of \citet{Lellouch2006} for the Jupiter SL9 impact, we assume that impact chemistry and the subsequent photolysis of \nhthree\ produce a CO-to-HCN ratio of 12. Assuming an initial CO mass load of $(2.1\pm0.4)\times10^{15}$~g, as proposed by \citet{Cavalie2010}, this corresponds to an initial nitrogen mass load in the form of HCN of $(8.7\pm1.7)\times10^{13}$~g. Comparing this with the nitrogen mass inferred from our observations, $(3.7\pm0.9)\times10^{10}$~g, implies that the HCN abundance has decreased by a factor of $(7.3^{+4.2}_{-2.6})\times10^3$. Given that \citet{Cavalie2010} dated the impact on Saturn $220\pm30$ years ago, this yields an e-folding time of $25\pm5$~years. This is longer than the one derived for Jupiter by \citet{Cavalie2023}, $\sim8$~years, but remains plausible, as photolysis, which is about four times less efficient on Saturn than on Jupiter, may still contribute to Jupiter's HCN decrease. Overall, given the current observational and theoretical uncertainties, it seems difficult to either favor or rule out a cometary impact as the origin of HCN in Saturn's stratosphere.

%%%%%%%%%%%%%%%%%%%%%%%%%%%%%%%%%%%%%%%%%%%%%%%%%%%%%%%%%%%%%%
\section{Conclusions}

Using ALMA observations, we have measured and mapped for the first time the abundance of HCN in Saturn's stratosphere. The HCN meridional distributions are uniform within the noise level at 0.1 and 1~hPa, with vmr mean values of $(7\pm2)\times10^{-11}$ at 0.1~hPa and $(4\pm1)\times10^{-12}$ at 1~hPa. This corresponds to a mean column density $(4\pm1)\times10^{12}$~cm$^{-2}$ between the 2 and 0.05-hPa pressure levels and a total mass of nitrogen atoms in the form of HCN of $(3.7\pm0.9)\times10^7$~kg.

From our observations, we draw four main conclusions.
\begin{enumerate}
    \item The steep vertical decrease in HCN abundance from 0.1~hPa to 1~hPa, observed at all sampled latitudes (20\degr S--60\degr N), suggests that a loss mechanism located in the lower stratosphere dominates the overall HCN loss. This finding is consistent with the similar conclusions reached for Jupiter \citep{Cavalie2023}, Titan \citep{Vinatier2007}, and recently, Uranus \citep{Cavalie2026}. A loss mechanism on aerosols is strongly suggested since the loss rate increases in regions of high particle number density, namely the polar regions of Jupiter and the lower stratosphere in all these objects.
    \item Approximately one year after the beginning of Cassini's proximal orbits, Saturn's atmosphere shows no detectable chemical response to the large equatorial influx of ring material detected by Cassini. The total mass of nitrogen atoms present as HCN in Saturn's atmosphere corresponds to only a few thousand seconds worth of the influx measured by INMS. This result strongly supports the conclusions of \citet{Moses2023} that the precipitating material does not significantly ablate in Saturn's atmosphere.
    \item We are unable to identify a unique source of nitrogen responsible for the observed HCN in Saturn's stratosphere. Instead, the measured mass load is consistent with multiple possible sources, including Enceladus, the \textit{Ring Rain}, comet impact, and indirect internal sources (\ntwo\ converted into HCN). The observed meridional uniformity of HCN tends to disfavor the scenario in which the HCN presence is due to a single localized source, such as Enceladus or \textit{Ring Rain}. It is quite possible that several sources contribute simultaneously to the HCN abundance.
    \item Given current models of the stratospheric \ntwo\ photo and ion-neutral chemistry \citep{Moses2023}, our observed HCN abundance is incompatible with a \ntwo\ abundance larger than $\sim10^{-8}$, an upper limit smaller than that predicted by some models of the internal disequilibrium species \citep{Wang2016}.
\end{enumerate}

Compared to other Giant Planets, the situation of Saturn remains particularly puzzling, with the rings equatorial influx, the largest---by far--- external source playing no role in the chemical structure of Saturn's stratosphere, while we cannot disentangle the effect of several smaller sources. To progress in understanding this situation, we propose some possible further observational and theoretical work:
\begin{itemize}
    \item Monitoring the temporal evolution of exogenous molecules (CO, HCN, \htwoo) could reveal a lagged chemical response to the equatorial ring influx. If the source were transient, active only during Cassini's proximal orbits, the incoming species and their chemical products may require time to diffuse downwards and reach pressure levels probed by infrared or submillimeter observations.
    \item Some sources, especially the \textit{ring rain}, are expected to be hemispherically asymmetric. Observing possible differences between the northern and southern abundances of exogenous species could help quantify the contribution of \textit{ring rain} to the overall external flux. The current season, Saturn's southern spring, is favorable for this observation.
    \item Detecting new species, like CS, CH$_3$CN, HC$_3$N, could help quantify the respective roles of impact processes, photochemistry and ion-neutral chemistry as performed by \citet{Moreno2017} for Neptune.
    \item Developing a theoretical model for HCN loss on aerosols would enable more reliable estimates of the mass flux needed to explain the observed HCN abundance.
    \item Applying coupled chemistry-transport models of Saturn's stratosphere, like that of \citet{Hue2015,Hue2016}, would help constrain the meridional distribution of the infalling flux that is consistent with the observed CO and HCN uniform meridional profiles, and help identify the relative contributions of different sources.
\end{itemize}

%%%%%%%%%%%%%%%%%%%%%%%%%%%%%%%%%%%%%%%%%%%%%%%%%%%%%%%%%%%%%%
\begin{acknowledgements}
This paper makes use of the following ALMA data: ADS/JAO.ALMA$\#$2017.1.00636.S. ALMA is a partnership of ESO (representing its member states), NSF (USA), and NINS (Japan), together with NRC (Canada), NSTC and ASIAA (Taiwan), and KASI (Republic of Korea), in cooperation with the Republic of Chile. The Joint ALMA Observatory is operated by ESO, AUI/NRAO, and NAOJ.

TF thanks Thomas Gautier and Julianne I. Moses for insightful discussions.
\end{acknowledgements}

% Please note that we have included the references below in
% order to compile the document, but we ask you to:

\bibliographystyle{aa}
\bibliography{ref}

\end{document}